\documentclass[final, runningheads]{llncs}
\usepackage[T1]{fontenc}
\usepackage{graphicx}
\usepackage{xcolor}

\usepackage[numbers]{natbib}

\usepackage{hyperref}
\usepackage{amsmath}
\newcommand{\itadata}{\footnotesize \textsl{Workshop Scientific HPC in the pre-Exascale era (part of ITADATA2024)}}
\usepackage{fancyhdr}
\fancypagestyle{empty}{\fancyhf{}\fancyhead[C]{\itadata}
  \fancyhead[R]{\includegraphics[width=.05\textwidth]{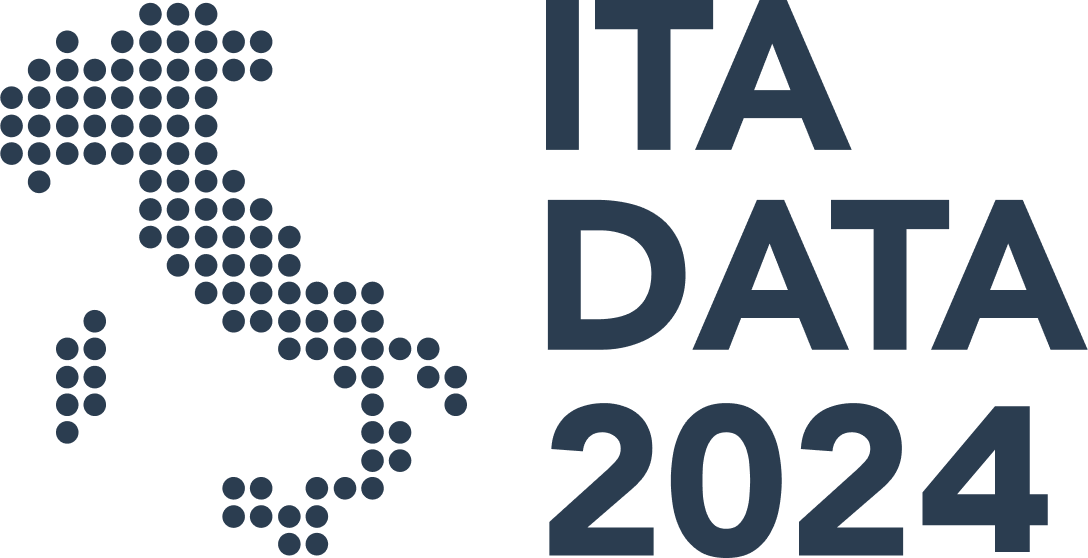}}}

\makeatletter
\begin{document}
\title{Green computing toward SKA era with RICK}
\author{Giovanni Lacopo\inst{1,2}\orcidID{0009-0008-0788-2516} \and
Claudio Gheller\inst{3,4}\orcidID{1111-2222-3333-4444} \and
Emanuele De Rubeis\inst{4,3}\orcidID{0000-0002-0428-2055} \and 
Pascal Jahan Elahi\inst{5}\orcidID{0000-0002-6154-7224} \and
Maciej Cytowski\inst{5} \and
Luca Tornatore\inst{2} \and
Giuliano Taffoni\inst{2} \and
Ugo Varetto\inst{5}
}
\authorrunning{G. Lacopo et al.}
\institute{Università degli studi di Trieste, via Alfonso Valerio 2, 34127 Trieste, Italy \and
Astronomical Observatory of Trieste INAF, via GB Tiepolo 11, 34143 Trieste, Italy
\email{info.oats@inaf.it} \and
Istituto di Radio Astronomia, INAF, Via P. Gobetti 101, 40129 Bologna, Italy \and
Dipartimento di Fisica e Astronomia, Universit\`{a} di Bologna, Via P. Gobetti 92/3, 40129 Bologna, Italy \and
Pawsey Supercomputing Centre, 1 Bryce Avenue, Kensington WA 6151, Australia
\email{{admin@pawsey.org.au}}}
\maketitle              \begin{abstract}
Square Kilometer Array is expected to generate hundreds of petabytes of data per year, two orders of magnitude more than current radio interferometers. Data processing at this scale necessitates advanced High Performance Computing (HPC) resources. However, modern HPC platforms consume up to tens of $MW$, i.e. megawatts, and energy-to-solution in algorithms will become of utmost importance in the next future. In this work we study the trade-off between energy-to-solution and time-to-solution of our \textbf{RICK} code (Radio Imaging Code Kernels), which is a novel approach to implement the $w$-stacking algorithm designed to run on state-of-the-art HPC systems. The code can run on heterogeneous systems exploiting the accelerators.
We did both single-node tests and multi-node tests with both CPU and GPU solutions, in order to study which one is the greenest and which one is the fastest. We then defined the \textbf{green productivity}, i.e. a quantity which relates energy-to-solution and time-to-solution in different code configurations compared to a reference one. Configurations with the highest green productivities are the most efficient ones. The tests have been run on the Setonix machine available at the Pawsey Supercomputing Research Centre (PSC) in Perth (WA), ranked as $28^{th}$ in Top500\footnote{\url{https://top500.org/lists/top500/list/2024/06/}} list, updated at June 2024.

\end{abstract}
\section{Introduction}
Radio astronomy is currently experiencing a sudden increase in the amount of data being gathered by radio-interferometers such as the Low Frequency Array \cite{2013A&A...556A...2V}, MeerKAT \cite{2016mks..confE...1J}, the Murchison Widefield Array \cite{2010rfim.workE..16M}, and the Australian Square Kilometre Array Pathfinder \cite{2007PASA...24..174J}. These instruments already produce petabytes of data annually. In the near future the data volume will increase by two orders of magnitude, with the Square Kilometre Array (SKA MID and LOW), anticipated to deliver hundreds of petabytes per year.

This will represent an exceptional technological challenge, since data processing will have to be performed exploiting the most advanced computational resources as those provided by High Performance Computing (HPC). Modern HPC systems are reaching performance through many-core and GPU based accelerated computing. Heterogeneous computing most probably will remain a common feature in emerging architectures. Benefitting from these novel hybrid architectures is non-trivial however, because of the challenges presented by mixed hardware computing and the increasing levels of architectural parallelism. New algorithms and numerical and computational solutions are required. Data reduction and imaging software tools will have to be adapted, and sometimes even completely re-designed, in order to efficiently utilize the available hardware.

The task becomes even more challenging if we consider that current pre-exascale HPC systems consume tens of $MW$, running under full workload, and it is easy to understand that this trend will become unsustainable in the near future when several exascale machines will be available. The energy efficiency of HPC platforms is measured as $\textup{energy efficiency} = GFlops/W$, i.e. billions of floating point operations per watt, and is classified in the Green500\footnote{\url{https://top500.org/lists/green500/list/2024/06/}} list.

We have focused on the development of a novel implementation of the $w$-stacking algorithm \cite{offringa-wsclean-2014} on state-of-the-art HPC systems, effectively exploiting heterogeneous architectures and with the goal to be sustainable in energy-to-solution. The $w$-stacking technique is one of the approaches for the imaging of radio-interferometric data, allowing the production of energy flux maps of the observed sky. Imaging is a computationally intensive step in the data processing pipeline \cite{1999ASPC..180.....T}, requiring a significant amount of memory and computing time. This is due to operations such as gridding, which involves resampling the observed data on a computational mesh, and Fast Fourier Transform (FFT), which transform back from Fourier to real space. For large fields of view Earth curvature effects cannot be ignored, leading to a fully three-dimensional problem: here the "$w$-term" correction needs to be introduced \cite{2008ISTSP...2..647C,offringa-wsclean-2014}. The gridding, FFT-transform, and $w$-correction steps are put together into the {\it $w$-stacking gridder}, that we have named RICK (Radio Imaging Code Kernels).

RICK combines both multi-core and accelerated solutions, as already discussed in \cite{gheller2023}, hereafter Paper I, and \cite{DERUBEIS2025100895}, hereafter Paper II. In Paper II, we presented a full GPU implementation of the code, in which data are loaded on the accelerators at first step and CPUs are utilized eventually just to write the final image. However, we stressed that distributed FFT on GPUs is not portable and currently works on NVIDIA GPUs only\footnote{\url{https://docs.nvidia.com/hpc-sdk/cufftmp/index.html}}. For all the other configurations FFT still runs on CPUs and we rely on the distributed FFTW\footnote{\url{https://fftw.org/doc/Distributed_002dmemory-FFTW-with-MPI.html}} or on its hybrid MPI+OpenMP version\footnote{\url{https://www.fftw.org/fftw3_doc/Combining-MPI-and-Threads.html}}. Energy efficiency of CPUs and GPUs has been explored extensively in many works, we cite \cite{qasaimeh2019comparing,zeng2021energy} as reference, which confirm that heterogeneous computing is currently the best trade-off in terms of energy-to-solution sustainability.

In this paper we study various CPU only and accelerated solutions in order to determine optimal configurations in terms of both time-to-solution and energy-to-solution for radio astronomy.\\
The paper is organized as follows: in Section 2 we will discuss the theory behind the code and its implementations, in Section 3 we will define the \textbf{green productivity} and its relevance for our code. Results will be presented and discussed in Section 4. Conclusions will be drawn in Section 5.

\section{The RICK code}
\subsection{Theory}
When large Fields-of-View (FoV) are observed at once, visibility data from non-coplanar interferometric radio telescopes cannot be accurately imaged with a two-dimensional Fourier transform only and the imaging algorithm needs to include the correction for the $w$-term~\cite{2008ISTSP...2..647C}. This term expresses the deviation of the array from a plane. A possible approach to account for the $w$-term is constituted by the $w$-stacking method \cite{offringa-wsclean-2014}, where the computational mesh has a third dimension in the $w$-direction and visibility data are mapped to the closest $w$-plane.
The correction is applied as:
\begin{equation}
\begin{split}
\frac{I(l,m)\left(w_{\max} - w_{\min}\right)}{\sqrt{1-l^2-m^2}} = &\int\limits_{w_{\min}}^{w_{\max}} e^{2\pi i w(\sqrt{1-l^2-m^2}-1)} \times 
\\
&\iint V(u,v,w)  e^{2\pi i \left(ul + vm\right)} du dv dw.
\label{eq:wstacking}
\end{split}
\end{equation}
with:
\begin{equation}
\begin{split}
V(u,v,w) = &\int\int \frac{I(l,m)}{\sqrt{1-l^2-m^2}} \times \\ 
         &e^{-2\pi i \left(ul + vm + w(\sqrt{1-l^2-m^2}-1)\right)} dl dm,
\label{eq:visI}
\end{split}
\end{equation}
where $u,v,w$ are baseline coordinates in the coordinate system of the antennas, $I$ is the spectral brightness and $l,m$ are the sky coordinates. For small FoV, the term $\sqrt{1-l^2-m^2}$ is close to one, and Equation \ref{eq:visI} is an ordinary two-dimensional Fourier transform, which, in order to speed-up the computation, is solved by using Fast Fourier Transforms (FFT). 

This procedure, involves mapping point-like visibility data onto a regular two-dimensional mesh, to discretize the ($u$,$v$) space. This is obtained by convolving the visibility data with a finite-size kernel, resulting in a continuous function which can then be FFT transformed.

Two main data structures characterise the memory request of the algorithm. The first is an unstructured dataset storing the ($u,v,w$) coordinates of the antennas array baselines at each measurement time. Each baseline has a number of associated visibility data, which is determined by the frequency bandwidth, the frequency resolution 
and the number of correlations. A further quantity, the weight, is also assigned to each measurement. Visibilities are distributed among multiple memories in time slices of equal length, which is not strictly necessary but makes results interpretation much easier. The second, is a Cartesian computational mesh of size $N_u \times N_v \times N_w$, where $N_u$, $N_v$ and $N_w$ are the number of cells in the three coordinate directions. The convolved visibilities and their FFT transformed counterpart are calculated on the mesh. Data defined on the Cartesian computational grid are split among multiple memories adopting a rectangular slab-like decomposition, assigning to each task a rectangular sector (or slab) of $N_{\rm{mesh}} = N_u \times N_w \times (N_v/N_{\rm{pu}})$ cells, where $N_{\rm{pu}}$ is the number of physical memories, i.e. MPI processes, adopted in the computation. The two data structures determine the memory request of the algorithm and they are evenly distributed among different physically disjoint memories. This decomposition could be changed to use pencil like decomposition but the default FFTW implementation's request forces us to utilize slab like decomposition. 

The code is composed of five primary algorithmic components, which are illustrated in Figure~\ref{fig:workflow}. These components are designed to support various levels of HPC implementations. The initial component handles the task of extracting observational data from binary files stored on the disc. The files are read in parallel, with each parallel task receiving an equal portion of the data. However, this is a time-ordered data distribution, and in a successive step data must be reorganised in a space-ordered distribution.

The following step performs the gridding of the visibility data. Gridding is done in successive rectangular sectors along the $v$-axis. The gridding procedure consists of two sub-steps. In the first sub-step an array is created for each sector, concatenating the data with $u$-$v$ coordinates inside the corresponding sector. 
The second sub-step is represented by the convolution with gridding kernel, which can be either a Gaussian or a Kaiser-Bessel~\cite{jackson_91} kernel:
\begin{equation}
    \tilde V(u_i,v_j,w_k) = \sum_{m \in {\rm measures}} V_m G((u_m,v_m,w_m),(u_i,v_j,w_k)),
\label{convolve}
\end{equation}
where $m$ is the $m$-th measurement, $(u_m,v_m,w_m)$ are its coordinates, $(u_i,v_j,w_k)$ is a computational grid point, $V_m$ is the measured visibility and $\tilde V$ is the visibility convolved on the mesh. Once boundary data among different sectors are coped with, this part of the computation is completely local, and each sector can be executed on multiple cores through multithreading or on the accelerator. Multithread execution requires proper management of the accumulation of the contributions of each measurement on the computational mesh, that leads to frequent race conditions. This is solved using atomic updates functions provided by both OpenMP and CUDA/HIP. This step is finalised by communication among different computing units, necessary to accomplish the calculation of gridded visibilities on each sector. This is a critical operation, that is described in detail in~\cite{gheller2023}. Steps 1 and 2 are repeated in a loop until all input data, split among frequencies or time slices, are processed. However, time slices and frequency slicing can have different resulting science outputs and are related to different science cases. For a continuum image, one will stack all the frequencies but for spectral imaging, for instance, particularly for searches for absorption lines one might stack time together. So this outlined processing stages are for a standard spectral or continuum cube with stacking of observations.

The third algorithmic component performs the FFT of the gridded data, producing the real space image. This is accomplished exploiting the FFTW library. Next, we apply the correction for the $w$-term
and reduce the $w$-planes to obtain the final image (step 4). The $w$-planes reduction calculation are MPI-process local with the frequent race conditions in multi-threaded computation handled using atomic operations. In the fifth and final step the resulting image is written in a file (in parallel) on the disk. If GPUs are used, this step requires the preliminary copy of the image from the device to the host memory However, this is not necessarily true when running on an hardware that has RDMA-enabled IO.

\subsection{Code implementations}
\begin{figure} [ht]
\begin{center}
\includegraphics[height=6cm]{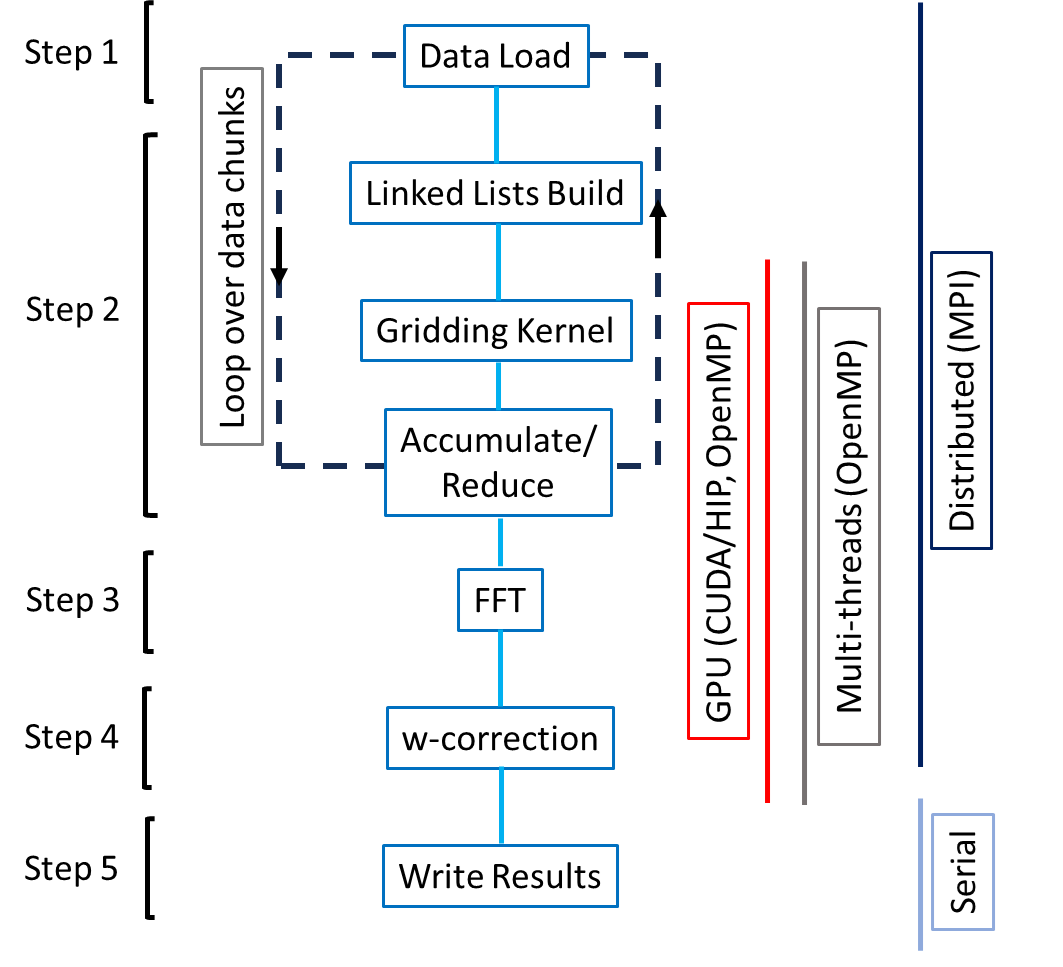}
\end{center}
\caption[example] {
\label{fig:workflow} 
RICK code workflow, highlighting the 5 main algorithmic components and the level of HPC enabling (MPI, multithreading, GPU) of each component.}
\end{figure}

The code is written in C, with C++ extensions just to add the CUDA/HIP support for GPUs. One of the goals of this code is the portability, in order to run on different architectures. In this section we will discuss the three implementations that have been tested in this work. We have a pure MPI implementation, an hybrid MPI+OpenMP implementation and a MPI+CUDA/HIP implementation to use GPUs.\\
As shown in Figure~\ref{fig:workflow} RICK performs the five algorithmic steps described above and can be run both in serial and in parallel. Of course the scale of SKA imaging necessitates massively parallel systems, so hereafter we will only consider the parallel implementations of the code.\\
Reading data from the filesystem is done in parallel, but the current implementation has also the possibility to handle datasets of any size even on a single node, because they can be easily managed by splitting stacked visibility data, i.e. large input data can contain visibilities from observations at different frequencies, into frequencies or time chunks. Instead of loading all the data at once, data coming from different frequencies are assigned to a chunk. Chunks can be loaded sequentially, as described in \cite{Gheller2024HPC}. This also enables to reserve an appropriate fraction of memory for
the computational mesh.\\
Data are initially distributed in time-log order, i.e. if input data come from 8 hours of observations and 8 MPI tasks are used, task 0 will read the first hour of observation, task 1 will read the second hour and so on. This does not require that each observation actually needs to be of the same length but interpretation of the time stacking does help if each observation is roughly the same. In order to reconstruct the sky image, each MPI task processes single sectors in the u-v plane, necessitating a  transformation from time-log order to space-log order. In time-log order input, each task has visibilities pertaining to all the other ones, and MPI communication is unavoidable. When visibilities in each sector are gridded, an MPI Reduce operation is performed on the grid with target the MPI task that owns the sector.

\subsubsection{MPI implementation}
In the MPI implementation one assigns an MPI task to each of the available CPU cores. This means that the five algorithmic steps performed by the code are fully distributed among all the MPI processes. There are as many sectors as MPI tasks and consequently the same number of reduce operation is needed, leading to a communication overhead (see Paper I and II).\\
Communication has been largely discussed in section 5 of Paper II, so a detailed treatment is beyond the scope of this paper. In this pure MPI case, we've used the standard MPI Reduce, available in every MPI implementation.

\subsubsection{Hybrid MPI/OpenMP implementation}
To mitigate the impact of MPI communication, 
it became essential to include in our code a hybrid parallelization with MPI and OpenMP. 
The hybrid implementation reduces the communication impact by diminishing the number of MPI Reduce calls and reduces the communication surface because MPI Reduce involves less MPI processes. Furthermore, we have designed a hybrid technique combining MPI and OpenMP and exploiting Non-Uniform Memory Access (NUMA) topology. It works as follows: as first, two MPI communicators are built: 
\begin{enumerate}
    \item An intra-node communicator, where only the MPI ranks pertaining to each computing node are included and are ranked from $0$ to $P-1$, with $P$ being the number of MPI ranks on every node. When multi-socket nodes or when multi-NUMA regions CPUs are available, MPI processes are automatically assigned to each NUMA region in a round-robin fashion.
    \item An inter-node communicator that groups all ranks $0$ (which are the node masters) in the intra-node communicators. 
\end{enumerate}
In communicator $1$, every rank knows about all other ranks in a given communicator so rank $i$ will communicate to rank $i+1$ in a ring with $P\rightarrow0$. Each MPI task spawns two threads. Thread 1 is in charge of calling the ring shared-memory reduce function involving all the other tasks in the intra-node communicator, while thread 0 manages the MPI communications among the nodes. It is important to say that although each task needs at least two threads, only task 0 is actually involved in inter-node MPI communications. So in the current implementation all the other tasks spawn \textit{idle} threads.\\
The reduce operation in one node is in shared-memory because MPI shared windows are used to create direct memory access channels between the MPI tasks in the same communicator. When the shared-memory reduce is done, 
task 0 (i.e the master) gathers the result: this is the total summation in case of a single node run, and it is a partial summation when there are multi-nodes. In this work hybrid tests have been performed on just one node. How the reduce is handled in multi-node cases is discussed in Section 3.3 of Paper II.

\subsubsection{GPU implementation}
To fully exploit the computing power of accelerators we have designed a GPU implementation of the code which performs three main steps of the algorithm directly on the accelerators: gridding, reduce, $w$-correction. Grid is allocated on the GPU memory at the beginning, whereas visibilities are loaded on the device in each sector, then the gridding function is turned to a gridding kernel which maps visibilities on the 2D mesh. After this step, the gridded data are entirely present in the GPU memory, such that direct GPU-GPU communication can be performed without transferring data back to the host. GPU-GPU communication is performed on AMD GPUs with the ROCm Collective Communication Library (RCCL\footnote{\url{https://rocm.docs.amd.com/projects/rccl/en/latest/}}), which implements the reduce operation as a ring intra-node, and an inter-node ring, where GPUs assigned to the MPI tasks communicate with Remote Direct Memory Access (RDMA) with GPUs in different nodes without passing through the CPUs ~\cite{Li2020,desensi2024exploringgputogpucommunicationinsights}. Fast Fourier Transform is still computed on the CPU because there is no available implementation for AMD GPUs yet.

\section{Green productivity}
The goal of this work is to find the best trade-off in the code's energy efficiency and performance when \textit{more computing resources/different configurations} are used. However we considered that although is easy to measure either when a code is faster or less energy consuming, it is not trivial to find a physical quantity that relates the energy consumption with the runtime in a meaningful way. This led us to the introduction of the \textbf{green productivity}. It has the following definition:
\begin{equation}
    GP = \frac{T_0/T_N}{\alpha E_N/E_0}
    \label{equation:eq}
\end{equation}
Here $T_0$ and $E_0$ are runtime and energy consumption of a reference configuration, respectively. Clearly, when the same code implementation is being tested and the only thing changing is the number of computing resources, the quantity at the numerator is simply the speedup in a strong/weak scaling test. $\alpha$ is a weight factor which changes depending on what we consider more important among performance and energy consumption. In this work we treated them with the same importance and chose $\alpha=1$.

\section{Results and discussion}
In this Section we present and discuss the different tests with the corresponding results. We focused on three different code's implementations, i.e. pure MPI, hybrid MPI+OpenMP and MPI+HIP. 
We relied on the SLURM (Simple Linux Utility for Resource Management) energy counters for the energy measurements in both CPU only and CPU+GPU tests.\\
With these counters, the energy consumption of the entire job is measured, apart from I/O and memory accesses, even though the energy consumed is still impacted by time spent doing IO. Here we have focused on the total energy consumption of the whole code. Other libraries, like PAPI\footnote{\url{https://github.com/icl-utk-edu/papi}}, permit the energy profiling of codes' functions with internal calls.

\subsection{Experimental setup}
Input data in our tests come from 8 hour observations from the LOFAR HBA Inner Station at different frequency channels, with each channel using $\sim 4.4GB$ of storage. We stacked two out of these datasets for the \textit{Single-node} tests and 18 of them for the \textit{Multi-node} tests. Details on memory occupancy are shown in Table \ref{table:datasets}.\\
The tests have been run on the Setonix-CPU and Setonix-GPU machines available at the Pawsey Supercomputing Research Centre (PSC) in Perth (WA). Setonix-GPU is ranked as $28^{th}$ in the June 2024 Top500 list and $10^{th}$ in the June 2024 Green500 list.\\
CPU partition is made of 1088 computing nodes equipped with 2 $\times$ AMD EPYC 7763 "Milan" 64 cores; GPU partition is made of 154 computing nodes equipped with 1 $\times$ AMD optimised 3rd Gen EPYC "Trento" 64 cores and 8 GCDs (from 4x "AMD MI250X" cards, each card with 2 GCDs), 128 GB HBM2e. CPU-GPU and GPU-GPU interconnections inside each node are guaranteed by the InfinityFabric technology. For node-node connections, both partitions have Slingshot interconnections. In all the tests the code has been compiled with \textbf{clang-16}\footnote{\url{https://releases.llvm.org/16.0.0/tools/clang/docs/ReleaseNotes.html}} and \textbf{MPICH}\footnote{\url{https://docs.nersc.gov/development/programming-models/mpi/cray-mpich/}} implementation has been utilized.

\begin{table}
\begin{center}
\centering \tabcolsep 2pt
\begin{tabular}{|l|l|l|l}
\hline
   & Single-node & Multi-node\\ 
\hline
N. visibilities (approx) & 1.08$\times$10$^9$ & 9.72$\times$10$^9$\\
Input data size (GB) & 8.8 & 78\\
$N_u \times N_v \times N_w$ & $4096^2\times$64 & $16384^2 \times 24$\\
Mesh size (GB)  & 43.24 & 194.11\\
\hline
\end{tabular}
\end{center}
\caption{Configuration and computational mesh used in the \textit{Single-node} and \textit{Multi-node} tests. The mesh size takes into account the total amount of memory required for the real and imaginary part depending on the size of the grid.}
\label{table:datasets}
\end{table}

\subsection{Single-node tests}
\begin{figure} [ht]
\begin{center}
\includegraphics[height=6cm]{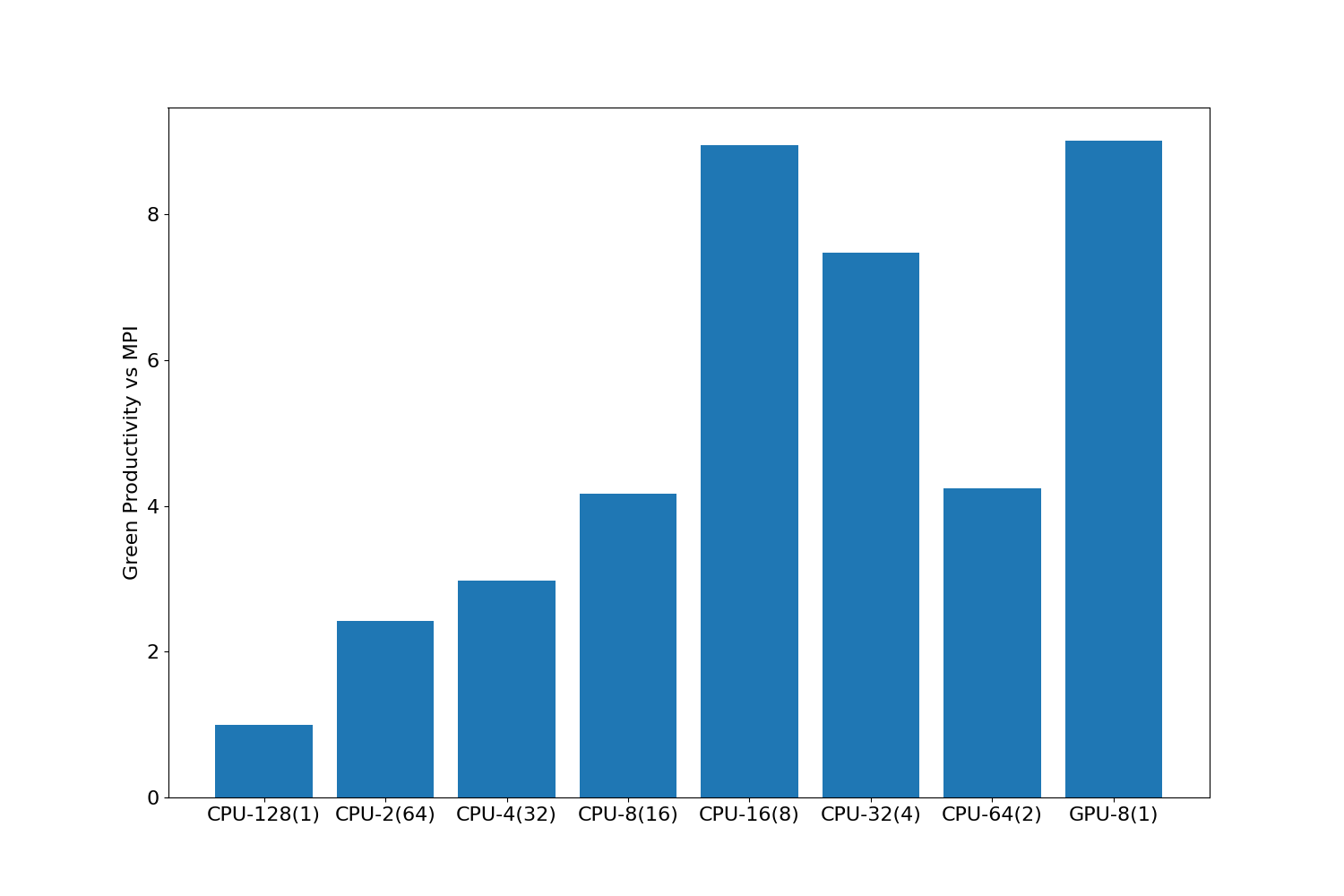}
\end{center}
\caption[example] {
Green productivity referred to the pure MPI run of the different hybrid MPI+OpenMP configurations and CPU+GPU.
\label{fig:single_node} 
}
\end{figure}

Here we show the results for several CPU hybrid MPI+OpenMP configurations and the CPU+GPU configuration referred to the pure MPI run. As we mentioned in Section 2, the pure MPI tests utilize the standard MPI Reduce function deployed by the MPI library, whereas MPI+OpenMP tests utilize our hybrid reduce implementation and GPU tests utilize RCCL reduce implementation. In Table \ref{table:reduce_diff}, we show the results for the pure MPI test, the best configuration for the hybrid MPI+OpenMP test and CPU+GPU test. Results are averages and standard deviations over four runs for each configuration. The hybrid implementation is $4$ times faster and $3.3$ times more green than the pure MPI run, while GPU implementation is $8.2$ times faster and $3$ times more green than pure MPI. The main difference is in the reduce operation, which is $20$ and $40$ times faster for the hybrid and GPU implementation compared to MPI, respectively. The performance difference in FFTW in GPU runs is due to the fact that it's still on the CPU and no OpenMP threadization is active for this test. Pure MPI tests are the fastest in gridding time, even when comparing them to GPU tests. This behaviour is due to the overhead related to HIP GPU memory management, and, more specifically, to repeated \textit{hipMalloc} and \textit{hipFree} calls that are implied by the iteration procedure through mesh sectors, as discussed in Paper II. However, in order to understand which configuration is actually the most efficient we show in Figure \ref{fig:single_node} the green productivity, where $T_0$ and $E_0$ in \eqref{equation:eq} are $T_{MPI}$ and $E_{MPI}$. We notice that the hybrid implementation with 16 MPI tasks spawning 8 OpenMP threads each and the CPU+GPU configuration turn out to have a similar green productivity. 

We can make the following considerations: $1)$ when the green productivity is almost the same it's up to the user whether to choose the fastest or the most green configuration; $2)$ it is important to add that, at least at the Setonix machine, asking for GPUs requires $8$ times more core hours. For this reason, when a problem does fit in a single node, we found that the best solution is the hybrid one.

\begin{table}
\begin{center}
\centering \tabcolsep 2pt
\begin{tabular}{|l|l|l|l|}
\hline
    & MPI & Best MPI/OMP & GPU\\ 
\hline
Energy (KJ) & $60.8375 \pm 2.6709$ & $18.3325 \pm 0.2175$ & $20.2325 \pm 0.3790$\\
Total time (sec) & $95.9533 \pm 0.1560$ & $24.4133 \pm 1.7793$ & $11.7309 \pm 0.0446$\\
Gridding time (sec) & $0.6624 \pm 0.0013$ & $1.5677 \pm 0.0183$ & $1.0925 \pm 0.0001$\\
Reduce time (sec) & $84.3082 \pm 0.1399$ & $4.3494 \pm 0.0054$ & $2.0065 \pm 0.0118$\\
FFTW time (sec) & $1.1941 \pm 0.0982$ & $1.8458 \pm 0.0561$ & $4.8540 \pm 0.0205$\\
$w$-corr. time (sec) & $0.4039 \pm 0.0007$ & $0.5201 \pm 0.0073$ & $0.1112 \pm 0.0003$\\
\hline
\end{tabular}
\end{center}
\caption{Total energy and runtimes for the relevant code portions in the MPI, MPI+OpenMP and GPU cases respectively. Here we report averages and standard deviations over four runs each.}
\label{table:reduce_diff}
\end{table}

\subsection{Multi-node tests}
This section will be devoted to the tests with the large input data and grid size involving many computing nodes. These are strong scalability tests, to understand the memory imprint of the code compared to the computational part. We will investigate energy-to-solution and time-to-solution of pure CPU and CPU+GPU configuration. In the end we will study the green productivity referred to the lowest resources solution for each configuration. 
\subsubsection{CPU tests}
Here we discuss the results for the CPU tests. We have chosen a combination of input data and grid size that doesn't fit in a single node, in order to include network traffic related to IO operations using the Lustre filesystem, which does impact on the reduce operation. We have done a strong scalability test with 2 nodes (256 MPI tasks), 4 nodes (512 MPI tasks), 8 nodes (1024 MPI tasks), 16 nodes (2048 MPI tasks), 32 nodes (4096 MPI tasks). We argued that multiple reduce calls are required by the code, especially in pure MPI runs. This means that the code is expected to be strongly memory-bound.\\
To study the memory imprint and the energy-to-solution in the code, we have run RICK under the same dataset and grid size configurations, but modifying at each step the CPU frequency. In particular, Setonix-CPU allows to select:

\begin{itemize}
    \item \textbf{Default frequency}: The frequency is established by the OS depending on the computing demand of the specific code function. It changes during the code execution but profiling reveals that it remains close to the highest CPU frequency achievable.
    \item \textbf{High frequency}: The highest frequency achievable from the CPU, which is $2.60 GHz$.
    \item \textbf{Medium frequency}: The CPU frequency is set to $2.00 GHz$.
    \item \textbf{Low frequency}: The CPU frequency is set to $1.50 GHz$. 
\end{itemize}

\begin{figure} [ht]
\begin{center}
\includegraphics[height=6cm]{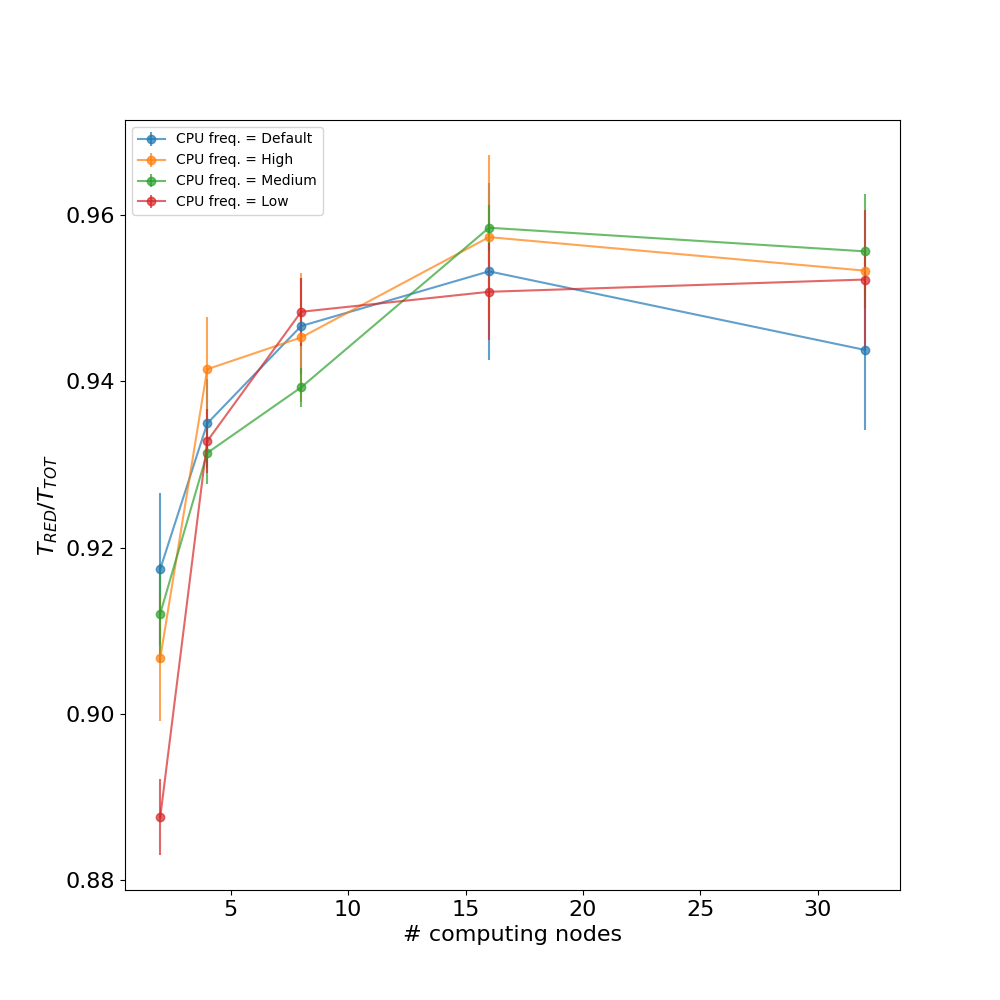}
\end{center}
\caption[example] {
Fraction of runtime spent in the reduce operation as a function of the number of computing nodes, for different CPU frequencies.
\label{fig:reducef} 
}
\end{figure}

We show in Figure \ref{fig:reducef} the fraction of runtime spent in the reduce operation as a function of the number of computing nodes, by changing the CPU frequency. 
We notice that this fraction increases with the number of nodes, eventually saturating around $\sim 95-96\%$ in all the configurations. In this case, when the reduce impact dominates the runtime, it makes sense to diminish the CPU frequency because the actual computational part is almost negligible. In this test, the pure MPI configuration has been used, and no MPI+OpenMP test is available. Indeed, the hybrid code implementation would have been diminished the communication surface and, as a consequence, the reduce runtime fraction in Figure \ref{fig:reducef}. 

The lack of the MPI+OpenMP test is due to the following shortcomings:

\begin{itemize}
    \item \textbf{Problem size and buffer limitations}: The hybrid reduce relies on the MPI Ireduce, i.e. the non-blocking MPI Reduce function, for node-node communications through the network. However, partial results are collected by node's master, which then performs the MPI communication with the other masters. If the problem size is too large the master rank cannot allocate such a huge buffer needed by the Ireduce implementation, resulting in a code crash. This has been a problem for the strong scalability test, especially when the number of computing nodes is low. Filling all the available cores with MPI allows each process to allocate smaller and smaller buffers when the problem size is fixed. Running the code with a small grid would have been outside the scope of this paper since no relevant scalability test could have been done and the arithmetic intensity of the code would have been reduced as well.
    \item \textbf{Lack of inter-node hybrid reduce}: The current implementation of the hybrid reduce is optimized for intra-node only. The implementation of a ring reduce among the masters of each node is under development, and avoids the bufferization problem discussed above by utilizing MPI shared windows and MPI Put/Get functions. The main difficulty happens again for large problem sizes, since the entire communication cannot be performed all at once and data need to be split in chunks to be shared one-by-one. This requires several MPI Win Flush calls, which currently introduce bottlenecks and sometimes to code hangs. In a future work we plan to fix this bugs and perform MPI+OpenMP CPU tests for node-node configurations as well.
\end{itemize}

\begin{figure} [ht]
\begin{center}
\includegraphics[height=6cm]{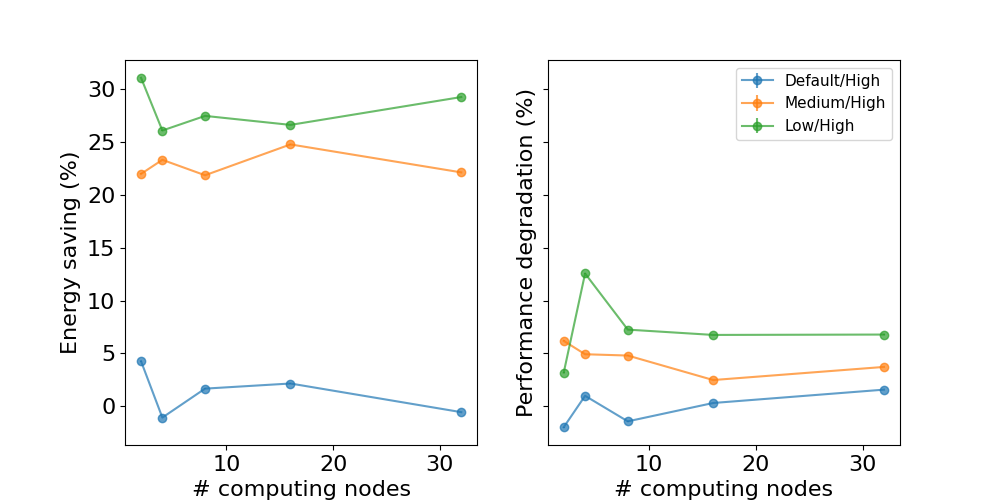}
\end{center}
\caption[example] {
Left: energy saving compared to the highest CPU frequency for default (blue), medium (orange), low (green) CPU frequencies as a function of computing nodes. Right: performance degradation of default, medium and low CPU frequencies compared to the highest CPU frequency as a function of computing nodes.
\label{fig:multi_node_save} 
}
\end{figure}

In the left panel of Figure \ref{fig:multi_node_save} we show the fraction of energy savings when we compare the default, medium and low frequencies to the highest frequency, as a function of the computing resources. We notice that default and high frequencies have approximately the same energy consumption, with small oscillations which anyway do not exceed $5\%$. It is more interesting what happens when we set the medium and low frequencies, because we reach up an energy saving of $25\%$ and $30\%$ when compared to the highest frequency, respectively. The right panel of Figure \ref{fig:multi_node_save} shows the performance degradation which does happen when we change the CPU frequency. When we set the default frequency, we notice that the code is slower by $2-4\%$ with some oscillations meaning that in certain code portions the OS sets up the default frequency to the maximum frequency. For the medium and low frequencies, we find a performance degradation of $4-5\%$ and $8-10\%$ on average, respectively. Thus we have around $5$ times more energy saving than performance degradation in percentage when we set the medium CPU frequency and around $3$ times when we set the lowest CPU frequency.

\subsubsection{GPU tests}
\begin{figure} [ht]
\begin{center}
\includegraphics[height=6cm]{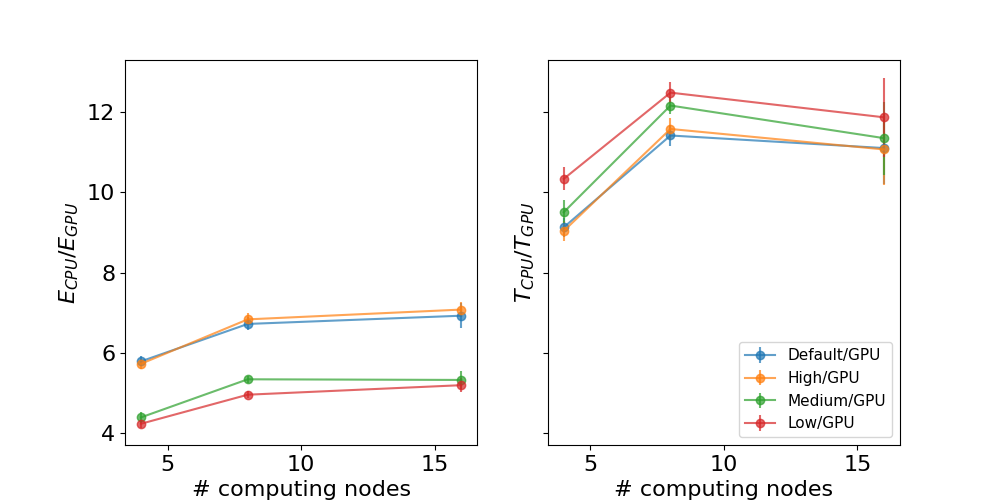}
\end{center}
\caption[example] {
Left: ratio between the energy in the pure CPU case and the energy in the CPU+GPU case, at different CPU frequencies, as a function of computing nodes. Right: ratio between the CPU runtime and CPU+GPU runtime, at different CPU frequencies, as a function of computing nodes. In the GPU tests both CPU and GPU frequencies are set by the OS to their default values.
\label{fig:gpu_energy_runtime} 
}
\end{figure}

We have done tests with 4, 8 and 16 GPU nodes, equipped with 32, 64 and 128 accelerators, respectively. In the following tests gridding, reduce and $w$-correction have been done on GPUs, while the FFT is still performed with the FFTW on CPUs. In the left panel of Figure \ref{fig:gpu_energy_runtime} we show the ratio between the energy consumed by the pure CPU tests and the energy consumed by the CPU+GPU tests, the latter being $6-7$ times more green than the former with high/default frequencies and $4-5$ times more green with medium/low frequencies. In the right panel of Figure \ref{fig:gpu_energy_runtime} we show the ratio between the pure CPU tests' runtime and CPU+GPU tests' runtime, the latter being faster by a factor $9-11$ for high/default frequencies and a factor $10-12$ for medium/low frequencies. For these \textit{Multi-node} tests GPUs turn out to be definitely the best choice because they're better in both energy saving and performance.

\subsubsection{Green productivity}
\begin{figure} [ht]
\begin{center}
\includegraphics[height=6cm]{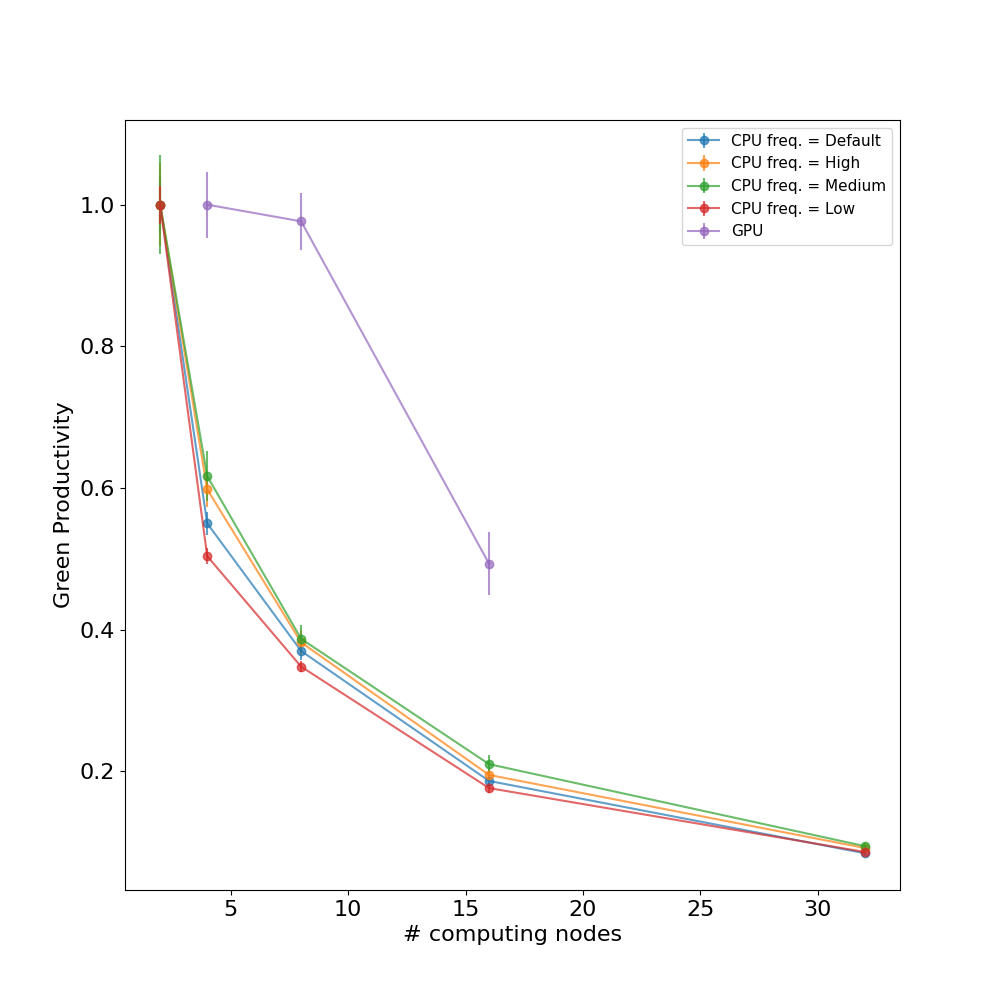}
\end{center}
\caption[example] {
Green productivity of CPU tests and CPU+GPU tests taking as reference the lowest node configuration for each different test, as a function of computing nodes.
\label{fig:green_prod} 
}
\end{figure}

In Figure \ref{fig:green_prod} we plot the green productivity of each specific configuration compared to its lowest node one, in this case $X_0 = X_2$ where $X$ can be both $T$ and $E$. Because the code is memory-bound, in strong scalability tests increasing the computing resources does not lead to a speedup. Sometimes the performance becomes worse because of the communication overhead due to the increase of reduce calls. The best configuration in the one with the highest green productivity, which in this case always corresponds to the lowest node configuration. However, in pure CPU cases green productivity drops much steeper even by moving from 2 to 4 nodes, whereas for GPUs it remains roughly constant when passing from 4 to 8 nodes. The sudden drop happening with 16 GPU nodes is explained by considering that when more and more accelerators are used by keeping the problem size constant, the gridding time stops scaling, again for overheads in HIP GPU memory management, as turned out in the discussion about \textit{Single-node} tests.

\section{Conclusions}
The results discussed in the previous section show that finding the configuration to achieve the best compromise between time-to-solution and energy-to-solution is not trivial. In RICK, it does depend on both data structures and the number of computing resources used.\\
Our results can be summarized as follows:
\begin{itemize}
    \item The usage of MPI and distributed GPU computing will be unavoidable when hundreds of petabytes of input data per year will be delivered, otherwise data processing in chunks is needed, raising large I/O overhead. 
    \item The accelerated version of the code is always faster than the pure CPU version, but when a problem fits in one node it is not the greenest solution.
    \item For large input data and very high resolutions, as it will be for SKA, many computing nodes will be needed to perform radio imaging, and GPUs become exceptionally faster and more green than CPUs, in particular due to the high difference in MPI communication.
    \item Green productivity is a measure to relate code performance and energy efficiency, and fine-tuning with the weighting factor is useful in order to focus more on one out of them.
\end{itemize}
To increase both the energy-to-solution and time-to-solution for the GPU implementation, our next step will be the inclusion of the distributed FFT version for AMD GPUs, which will avoid unnecessary memory movements back and forth from device to host, after the reduce operation, and then from host to device to perform $w$-correction at the end. 

\section*{Acknowledgements}
This paper is supported by the Fondazione ICSC, Spoke 3 Astrophysics and Cosmos Observations. National Recovery and Resilience Plan (Piano Nazionale di Ripresa e Resilienza, PNRR) Project ID CN\_00000013 "Italian Research Center for High-Performance Computing, Big Data and Quantum Computing" funded by MUR Missione 4 Componente 2 Investimento 1.4: Potenziamento strutture di ricerca e creazione di "campioni nazionali di R\&S (M4C2-19)" - Next Generation EU (NGEU), and it's also supported by (Programma Operativo Nazionale, PON), "Tematiche di Ricerca Green e dell'Innovazione". We acknowledge the Pawsey Supercomputing Research Centre for the availability of high performance computing resources, support and collaborations. The data to perform all the tests have been kindly provided by the LOFAR project LC14\_018 , PI F. Vazza.

\bibliographystyle{splncs04} \bibliography{franco,franco2,bib_add}

\end{document}